\documentclass[superscriptaddress,reprint,amsmath,amssymb,aps,prl,twocolumn,a4paper,floatfix]{revtex4-1}

\usepackage{amsmath,gensymb,textcomp,bm,dcolumn,eurosym,array,tabu,multirow,nicefrac,color,subfigure,graphicx,upgreek}
\usepackage[colorlinks, linkcolor=blue, citecolor=blue, urlcolor=blue, breaklinks=true]{hyperref}
\usepackage{amsmath,epsfig,amssymb,subfigure,bm,dsfont}
\usepackage{txfonts,times} 

\newcommand{\ket}[1]{\lvert #1 \rangle}

\usepackage[subsectionbib]{bibunits}
\usepackage{graphicx}
\usepackage{bm,MnSymbol}
\usepackage{units}
\usepackage{psfrag}
\usepackage{float}
\usepackage{lipsum}
\usepackage{color}

\begin{document}
\title{Dynamical control of quantum photon-photon interaction with phase change material}

\author{Chaojie Wang}
\affiliation{Department of Physics, Xiamen University, Xiamen 361005, China}
\author{Xutong Li}
\affiliation{Department of Physics, Xiamen University, Xiamen 361005, China}
\author{Xiuyi Ma}
\affiliation{Department of Physics, Xiamen University, Xiamen 361005, China}
\author{Yuning Zhang}
\affiliation{Department of Physics, Xiamen University, Xiamen 361005, China}
\author{Meng Wu}
\affiliation{Department of Physics, Xiamen University, Xiamen 361005, China}
\author{Weifang Lu}
\affiliation{Department of Physics, Xiamen University, Xiamen 361005, China}
\author{Yuanyuan Chen}
\email{chenyy@xmu.edu.cn}
\affiliation{Department of Physics, Xiamen University, Xiamen 361005, China}
\affiliation{School of Electronic and Optical Engineering, Nanjing University of Science and Technology, Nanjing 210094, China}
\author{Xiubao Sui}
\affiliation{School of Electronic and Optical Engineering, Nanjing University of Science and Technology, Nanjing 210094, China}
\author{Lixiang Chen}
\email{chenlx@xmu.edu.cn}
\affiliation{Department of Physics, Xiamen University, Xiamen 361005, China}

\begin{abstract}
Quantum interference can produce a pivotal effective photon-photon interaction, enabling the exploration of various quantum information technologies that beyond the possibilities of classical physics. While such an effective interaction is fundamentally limited to the bosonic nature of photons and the restricted phase responses from commonly used unitary optical elements, loss-induced nonunitary operation provides an alternative degree of freedom to control the quantum interference. Here, we propose and experimentally demonstrate a concise yet powerful tool to unravel fundamental features of quantum interference based on the phase change material vanadium dioxide. Since the insulator-metal transition in an elaborate vanadium dioxide thin film can create any desired particle exchange phase response, we show its tunability over the effective photon-photon interaction between paired photons that are entangled in the symmetric and anti-symmetric forms, which may introduce sophisticated nonunitary operations and functionalities into programmable optical platforms. These results provide an alternative approach to investigate the quantum light-matter interaction, and facilitate the use of quantum interference for various quantum information processing tasks such as quantum simulation and quantum computation.
\end{abstract}
\maketitle

Photonic quantum information processing provides revolutionary tools for various areas including secure cryptography \cite{gisin2002quantum,pirandola2015advances}, quantum sensing \cite{degen2017quantum}, quantum computing \cite{justin2013boson}, to name but a few. In these quantum technologies, the exploitation of quantum interference promises to enhance the performance of many quantum information processing tasks that beyond the possibilities of classical physics \cite{pan2012multiphoton}, making them compelling for enhancing quantum channel capacities \cite{barreiro2008beating,hu2018beating}, improving noise resilience \cite{bonato2008even,okano2013dispersion}, and even speeding up certain tasks in photonic quantum computation \cite{knill2001scheme,gao2018programmable}. Hong-Ou-Mandel (HOM) interference is a prototypical example of such a quantum phenomenon that lacks any counterpart in classical optics \cite{hong1987measurement}. It states the fact that identical photons that arrive simultaneously on different input ports of a beam splitter would bunch into a common output port as a consequence of their bosonic nature and the conventional unitary beam splitters \cite{hong1987measurement}. However, there are some restrictions on the transformation matrix that merely uses a unitary transformation.

Since it may be impossible to solve the well-known non-deterministic polynomial (NP)-complete problem only using unitary operators, nonunitary transformations have the potential to provide a solution for these problems, such as the realization of Abrams-Lioyd's gate \cite{abrams1998nonlinear} and a more efficient Fredkin gate \cite{li2022quantum}. In the context of some open physical systems with lossy time-evolution, nonunitary operators are necessary in modeling their interaction with the environment \cite{mostafazadeh2002pseudo,xie2024probabilistic}. Additionally, as quantum entanglement is fragile and susceptible to decoherence, selective filtering of photonic quantum entanglement based on nonunitary optics is developed to mitigate interactions between the quantum system and its deleterious environment \cite{mahmoud2025selective}. Boson sampling and quantum walks can be upgraded to incorporate fermionic or anyonic behaviors that can be directly mimicked by anomalous quantum interference based on nonunitary transformation \cite{tillmann2013experimental,sansoni2012two}.

To tackle the experimental implementations of these non-unitary operations, the lossy beam splitters based on graphene \cite{roger2016coherent,hong2024loss,thongrattanasiri2012complete}, metasurfaces \cite{li2021non}, surface plasmon polaritons \cite{benjamin2017science}, Cr-SiN-Cr thin films \cite{Vetlugin2022anti} and fully-integrated waveguides with engineered dissipation \cite{ehrhardt2022observation,klauck2019observation} have been widely explored to exhibit many fascinating phenomena. Moreover, dynamical and continuous control of quantum photon-photon interactions is essential for programmable photonic platforms. This problem has been merely investigated by using a nonunitary metasurface, wherein the unique anisotropic phase response is used to create two extreme eigen-operations and enable the continuous control over the effective interaction \cite{li2021non}. Thus, there is relatively little research devoted to the dynamical and continuous control of quantum photon-photon interaction, in particular providing innovated materials for integrated quantum photonics.

Phase change materials that feature various structural and electronic transition become concise yet efficient methods for realizing dynamically tunable optical devices \cite{lencer2008map}. Vanadium dioxide (VO$_2$) has an insulator-metal transition process that can be controlled by external stimuli including thermal, optical, electrical, electrochemical, mechanical, and magnetic approaches, corresponding to structural variation from the monoclinic phase to the rutile one \cite{shi2019recent,liu2018recent}. Due to this insulator-metal transition, VO$_2$ thin films show great difference in the electrical and optical properties, which enable numerous tunable optical tasks such as smart windows \cite{jaewoo2013suppression,gao2012nanoceramic}, memory metamaterials \cite{driscoll2009memory}, engineered dissipation in non-hermitian system \cite{harrington2022engineered}, optical switch \cite{markov2015optically}, switchable holography and digital encryption \cite{liu2025phase}.

\begin{figure*}[!htbp]
\centering
\includegraphics[width=1\linewidth]{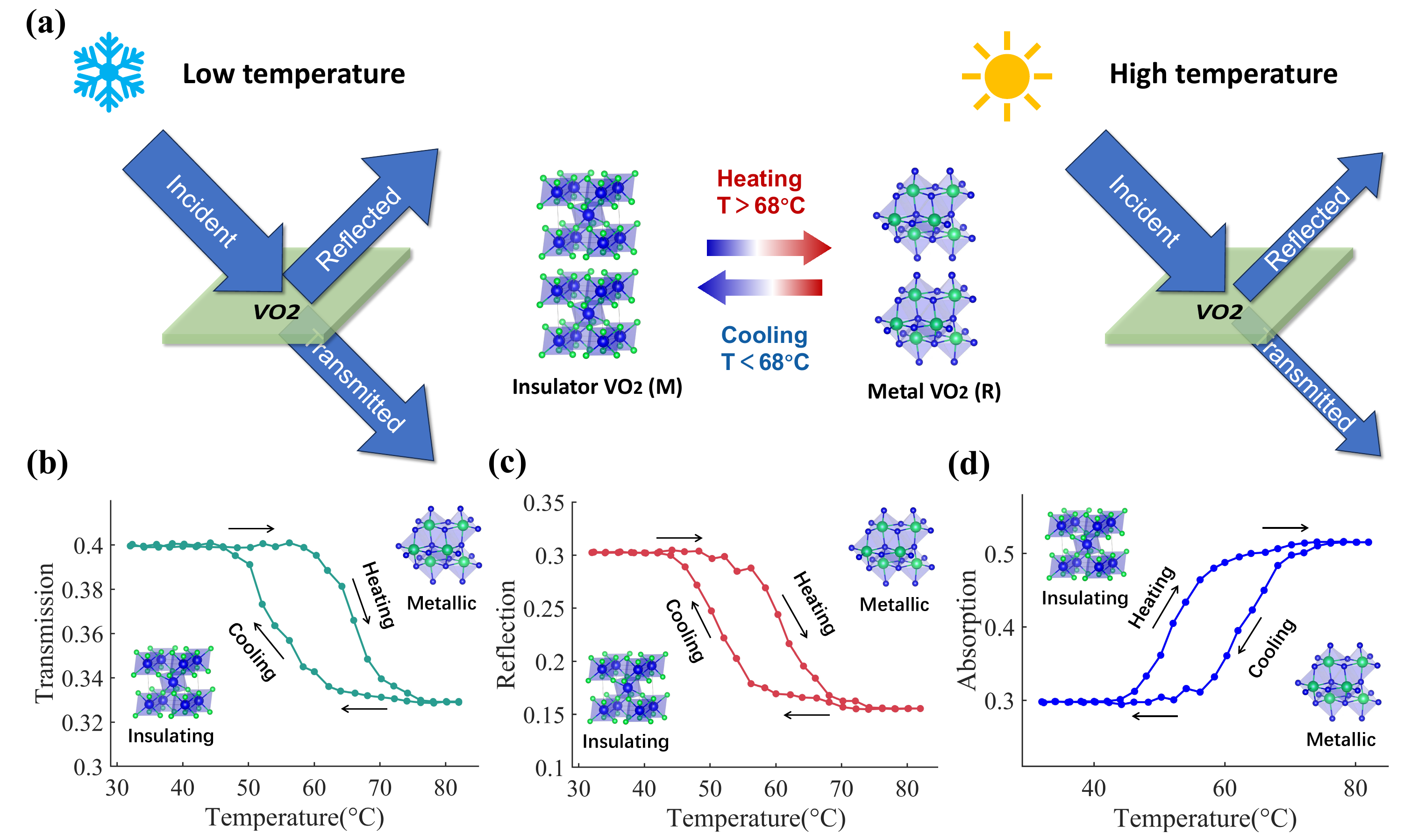}
\vspace{-2mm}
\caption{The optical property of VO$_2$ thin film can be tuned by external thermal stimuli, where (a) most photons are transmitted and reflected, and only a small amount of photons are absorbed at low temperature, but (b) only a small amount of photons are transmitted and reflected, and more photons are absorbed at high temperature. Its critical temperature for the insulator-metal transition is near 68$^\degree$C. (b-d) The measured transmission, reflection, and absorption probabilities of the VO$_2$ thin film with a thickness of 75nm that is used in our experiment.}
\label{figure_1}
\end{figure*}
Here, we leverage the distinctive properties of VO$_2$ thin films to realize the dynamical control of (non)unitary operation in quantum optics. To confirm its intriguing particle exchange statistics, an elaborate VO$_2$ thin film is used as the lossy beam splitter in HOM interference such that the coalescence and anti-coalescence of photons are dynamically and continuously controlled. Since VO$_2$ would behave as an insulator-metal-transition around a thermal temperature of 68$^\degree$C, we use an electrically driven ITO heater as the external stimuli to harness its loss, leading to the transformation between unitary and nonunitary operations. Additionally, tunable absorption of anti-symmetric quantum entangled photons in VO$_2$ thin film may provide an alternative approach for isolating a desired and steady entanglement state within a bosonic subspace, thereby providing a highly versatile linear mechanism for state selection through photon-photon interaction.

This work validates a promising way to use the phase change material in quantum physics, and sets the stage for advanced quantum technologies to be developed on programmable photonic platforms. We envisage that incorporating such phase change materials into the task of quantum information processing would dramatically enrich the available quantum operations and quantum functionalities in material, chemistry, biology, and photonic integrated circuits.

Vanadium dioxide is a technologically important metal oxide, owing to its remarkable change in insulator-to-metal transition at an easily accessible room temperature. Since its transmission and reflection contrast becomes maximized at the visible and MIR wavelengths, VO$_2$ provides great benefits for numerous applications such as actuators and thermochromic smart windows. Strikingly, we find that VO$_2$ thin film can behave as a beam splitter at the center wavelength of 810nm, namely the incident light would be transmitted, reflected, and absorbed as shown in Fig. \ref{figure_1}(a). In particular, its absorption coefficient is intimately related to the heating temperature. Figures \ref{figure_1}(b-d) present the measured transmission, reflection, and absorption probabilities of a VO$_2$ thin film with thickness of 75nm as a function of its thermal temperature, and this sample is also used in our HOM interference experiments. It is revealed that the absorption probability can range from 0.3 to 0.52, which is sufficient to introduce a tunable particle exchange phase and observe its corresponding transformation from unitary to nonunitary operations (Supplementary Note 1). In the meantime, there are still certain probabilities to obtain the transmitted and reflected photons such that the HOM interference experiment can be realized.

\begin{figure*}[!htbp]
\centering
\includegraphics[width=1\linewidth]{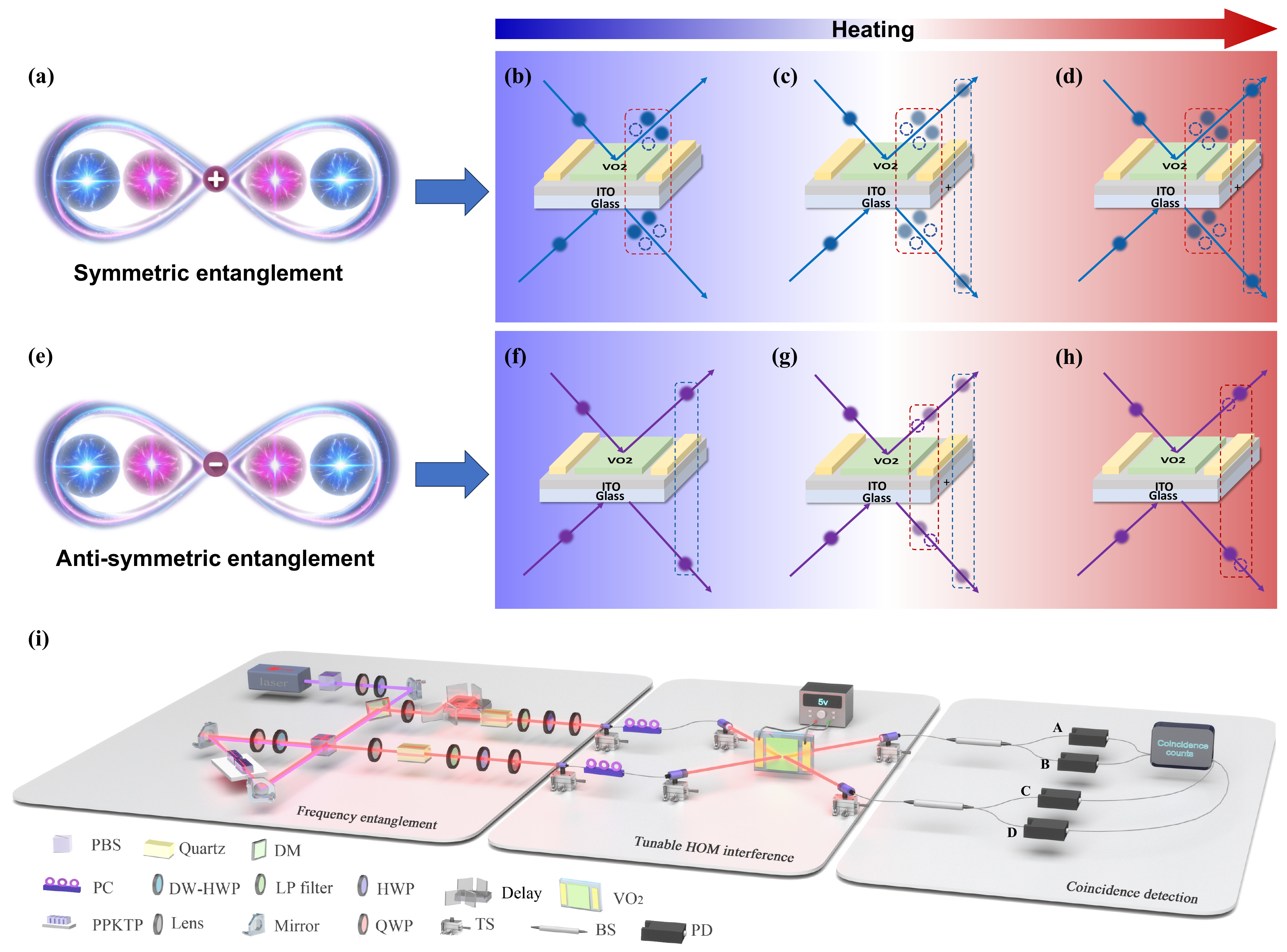}
\vspace{-2mm}
\caption{(a,e) The input frequency entanglement is prepared in the symmetric and anti-symmetric form that can be expressed as $\ket{\Psi}_{in}=(\ket{\omega_1\omega_2}\pm\ket{\omega_2\omega_1})/\sqrt{2}$. (b-d,f-h) Schematics of the distinct HOM interference output states (as illustrated by the photons (solid blue
balls) or no-photons (dotted blue circles) and associated dotted rectangles) based on insulator-metal-transition VO$_2$ thin film at zero delay when the heating temperature is 40$^\degree$C,65.5$^\degree$C, 80$^\degree$C, respectively. It corresponds to (b,f) the conventional bunching (anti-bunching) state for symmetric (anti-symmetric) and bosonic interaction, (c,g) one of the intermediate states and anyon-like interaction, and (d,h) the anomalous anti-bunching (bunching) state for symmetric (anti-symmetric) and fermion-like interaction. (i) The experimental setup of HOM interferometry. PBS, polarizing beam splitter; Quartz, compensation crystal; DM, dichromic mirror; PC, polarizing controller; DW-HWP, dual-wavelength half-wave plate; LP filter, long-pass filter; PPKTP, periodically poled potassium titanyl phosphate crystal; QWP, quarter-wave plate; TS, translation stage; BS: in-fiber beam splitter; PD, single photon detector.}
\label{figure_2}
\end{figure*}
Let us consider a typical configuration of HOM interferometer, where paired photons are incident on a VO$_2$ beam splitter from opposite input ports as shown in Fig. \ref{figure_2}(a-h). Without generality, the reflection, transmission, and loss coefficients of a beam splitter are defined as $r=|r|\exp(i\phi_r)$, $t=|t|\exp(i\phi_t)$, and $f$ by following the noisy operator theory \cite{barnett1998quantum}. Thus, the output states can be derived from the transformation of input states as
\begin{equation}
\begin{split}\label{bs}
\hat{a}_{out}=t\hat{a}_{in}+r\hat{b}_{in}+f_{a}\hat{n}_{a},\\
\hat{b}_{out}=t\hat{b}_{in}+r\hat{a}_{in}+f_{b}\hat{n}_{b},
\end{split}
\end{equation}
where the noise operator $\hat{n}_{a,b}$ may exhibit loss or gain to preserve the commutators of the observable outputs for a generic beam splitter. Since the physical constraint of energy conservation imposes restrictions on these parameters, the phase response is directly limited by the reflection and transmission coefficients by the inequality as 
\begin{equation}
\begin{split}
|\cos\phi_{rt}|\leq\frac{1-|t|^2-|r|^2}{2|t||r|},
\end{split}
\end{equation}
where $\phi_{rt}=\phi_r-\phi_t$ represents the difference between the phase response of the transmitted and reflected photons \cite{uppu2016quantum}. In a lossless and balanced beam splitter as $|r|^2=|t|^2=0.5$, the corresponding phase response becomes $\phi_{rt}=k\pi/2$ ($k$ is an integer) such that $r=\pm it$. In a lossy and balanced beam splitter as $|r|^2=|t|^2=0.25$, the corresponding phase response becomes $\phi_{rt}=k\pi$ such that $r=\pm t$. Thus, it is obvious that the loss engineering allows for the dynamical tuning of this particle exchange phase (Supplementary Note 1). 
\begin{figure*}[htbp]
\centering
\includegraphics[width=1\linewidth]{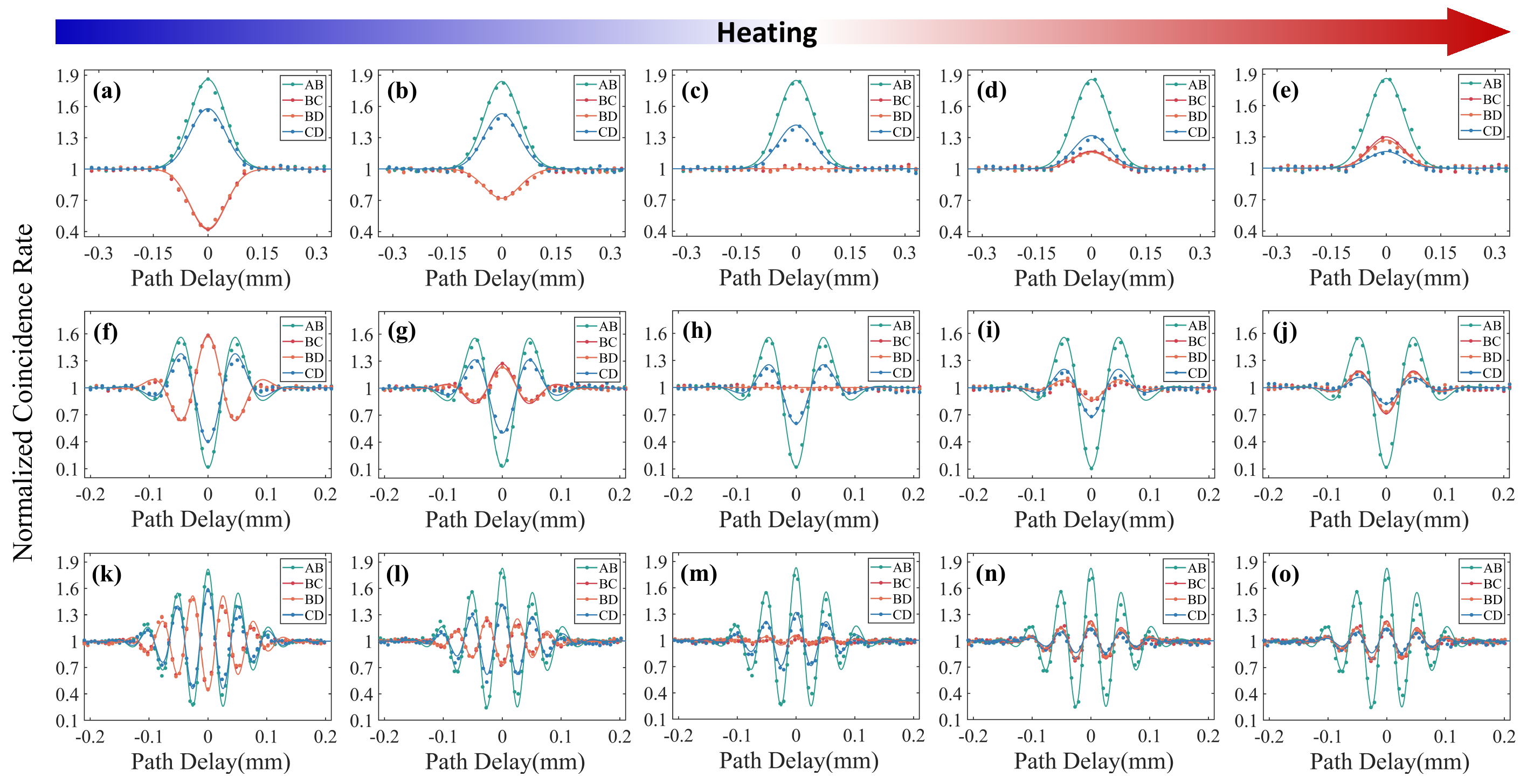}
\vspace{-2mm}
\caption{Experimental measurement of HOM interference patterns when heating the VO$_2$ thin films. To confirm the feasibility of controlling HOM interference by harnessing the symmetric properties of incident entanglement, symmetric entanglement with degenerate wavelengths (a-e), anti-symmetric entanglement with non-degenerate wavelengths (f-j) and symmetric entanglement with non-degenerate wavelengths (k-o) are used in our experiment. Before heating the VO$_2$ thin films, it behaves as a lossless beam splitter that works as a unitary operation, and thus the conventional HOM interference patterns corresponding to bosonic interaction are observed (a,f,k). Figures (b-d.g-i,l-n) that present the HOM interference patterns of anyon-like interaction are considered as the intermediate states. After heating the VO$_2$ thin films, it becomes a lossy beam splitter with absorption probability higher than 50\% such that it works as a non-unitary operation, and thus the anomalous HOM interference patterns corresponding to fermion-like interaction are observed (e,j,o).}
\label{figure_3}
\end{figure*}

\begin{figure*}[htbp]
\centering
\includegraphics[width=1\linewidth]{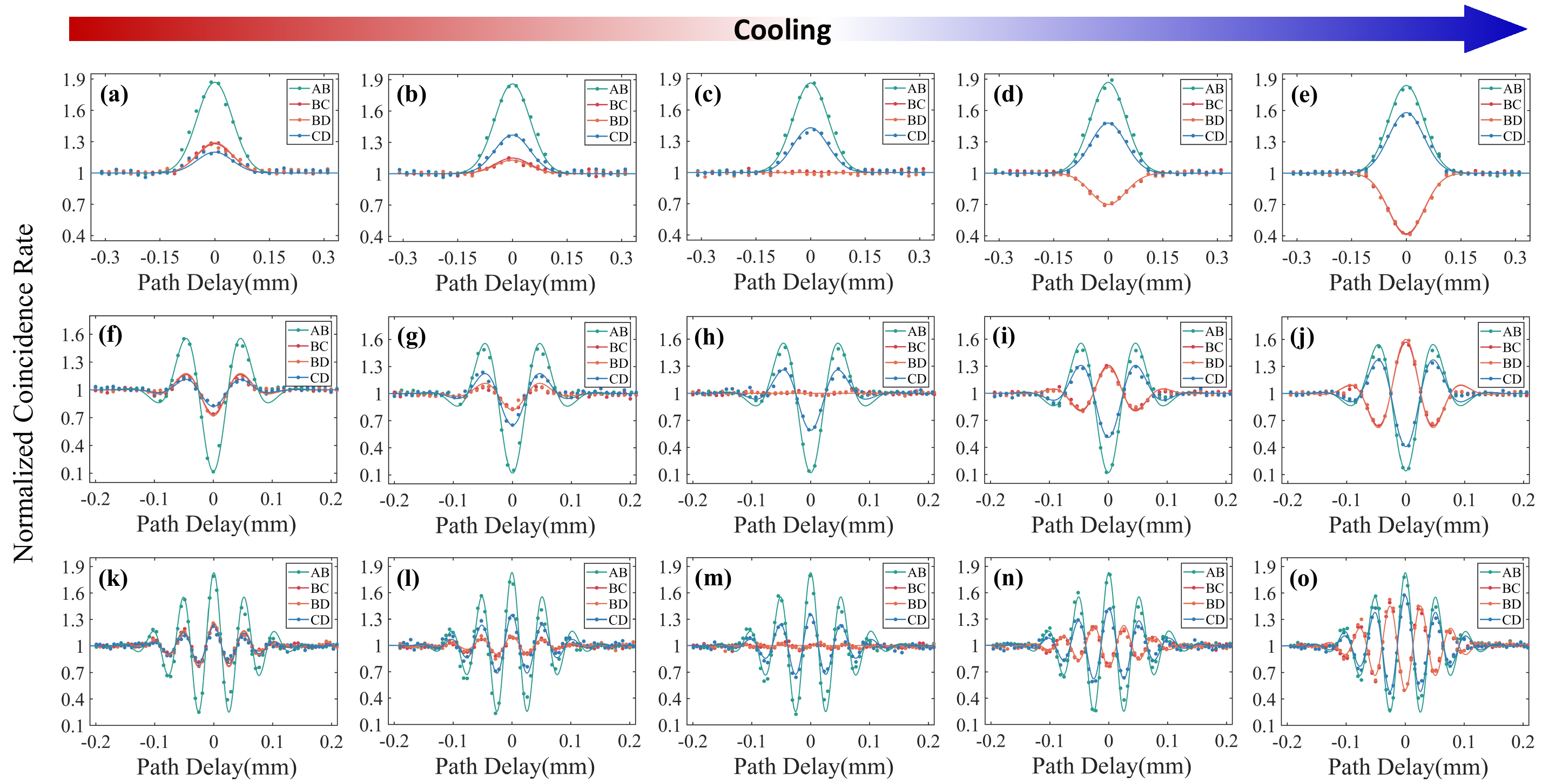}
\vspace{-2mm}
\caption{Experimental measurement of HOM interference patterns when cooling the VO$_2$ thin film. In analogy to Fig. \ref{figure_3}, the VO$_2$ thin film can change from lossy beam splitter (non-unitary operation) back to lossless beam splitter (unitary operation). This confirms that this VO$_2$ phase change material can be controlled with thermal temperature in a volatile and reversible fashion.}
\label{figure_4}
\end{figure*}
Here, we use the phase change material VO$_2$ thin film as the tunable beam splitter (Supplementary Note 3). To confirm this non-trivial phase response between the transmission and reflection coefficients, it is necessary to use two-photon quantum interference for the direct observation of particle exchange phases. We consider that paired photons entangled in the form of $\ket{\Psi}_{in}=(\ket{\omega_1\omega_2}+\exp(i\phi_\omega)\ket{\omega_2\omega_1})/\sqrt{2}$ are coherently superimposed on the VO$_2$ beam splitter \cite{chen2019hong}. By transmitting them through a beam splitter as depicted in \eqref{bs}, the output quantum state becomes
\begin{equation}
\label{bstransform}
\begin{split}
\ket{\Psi}_{out}\propto&(r^2+\exp(i\phi_\omega)t^2)(\ket{\omega_1^a\omega_2^b}+\ket{\omega_2^a\omega_1^b})\\
&+(rt+\exp(i\phi_\omega)rt)(\ket{\omega_1^b\omega_2^b}+\ket{\omega_1^a\omega_2^a}),
\end{split}
\end{equation}
where we neglect the events when one or two photons are absorbed. Quantitatively, the normalized coincidence probability detected at the identical and opposite output ports are expressed as
\begin{equation}\label{coincidence}
\begin{split}
&P_{(1,1)}(\tau)=|t|^4+|r|^4+2|t|^2|r|^2\cos(\Delta \tau +2\phi_{rt}+\phi_{\omega})e^{-\sigma^2\tau^2},\\
&P_{(2,0)}(\tau)=P_{(0,2)}(\tau)=2|r|^2|t|^2[1+\cos(\Delta \tau +\phi_{\omega})e^{-\sigma^2\tau^2}],
\end{split}
\end{equation}
where the subscript (1,1) ((2,0) and (0,2)) denotes the coincident events that paired photons are anti-bunched (bunched) into opposite (identical) output ports (Supplementary Note 2), $\Delta=|\omega_1-\omega_2|$ is the difference frequency of two well-separated center frequency bins, $\tau$ is the relative time delay introduced in the imbalanced HOM interferometry, $\sigma$ is the RMS bandwidth of single photons \cite{chen2019hong,ramelow2009discrete}. Thus, the tunable phase response of (non)unitary operations induced from the loss engineering in the VO$_2$ thin film can continuously transform the HOM interference dip (peak) to peak (dip), namely it can continuously transform the conventional coalescence of bosons to the anti-coalescence of bosons that allows for mimicking the fermionic or anyonic behaviors. As shown in Fig. \ref{figure_2}(a-h), we classify the input entanglement into two sets as symmetric entanglement ($\phi_\omega=0$) and anti-symmetric entanglement ($\phi_\omega=\pi$). Backed by the theoretical prediction in \eqref{coincidence}, the bunching probability is independent of the phase shift of nonunitary operations, which behaves as the phenomena that there are always a certain amount of symmetric entangled photons appearing in identical output ports as shown in Fig. \ref{figure_2}(b-d), and there are no anti-symmetric entangled photons appearing in identical output ports as shown in Fig. \ref{figure_2}(f-h). On the contrary, the anti-bunching probability is determined by the phase shift of nonunitary operations, which behaves as the transformation of symmetric entangled photons from disappearing to appearing in opposite output ports as shown in Fig. \ref{figure_2}(b-d), and the transformation of anti-symmetric entangled photons from appearing to disappearing in opposite output ports as shown in Fig. \ref{figure_2}(f-h).

The experimental implementation is shown in Fig. \ref{figure_2}(i), where two cascaded nonlinear crystals are used to create paired photons through the spontaneous parametric down conversion (SPDC) process, and their optical axes are mutually orthogonal for preparing tunable frequency entanglement (Methods and Supplementary Note 4). By adjusting its phase matching temperature, the frequency detunings $\Delta$ and the relative phases $\phi_\omega$ can be continuously tuned from $\Delta$=0 THz at 23.5$^\degree$C (symmetric entanglement with degenerate wavelength), to $\Delta$=2.95 THz and $\phi_\omega=\pi$ at 45$^\degree$C (anti-symmetric entanglement with nondegenerate wavelengths), and $\Delta$=5.85 THz and $\phi_\omega=0$ at 60$^\degree$C (symmetric entanglement with non-degenerate wavelengths). Subsequently, these paired photons are directed into a VO$_2$ beam splitter with a thickness of 75nm that is deposited on a 2mm-thick sapphire from different input ports. For harnessing the loss coefficient in this VO$_2$ thin film, an electrically driven ITO heater with a temperature sensor is used in our experiment. To extract the specific information about the coincidence counts arising from the paired photons that are bunched into the identical output mode, a 50:50 fiber beam splitter is placed at individual output port of HOM interferometry, which are identified as four single photon detectors labeled by A, B, C, and D. As a result, these coincidence events can be analyzed from different configurations of two random detectors (i.e., AB, AC, AD, BC, BD, and CD). Thereinto, the coincidence events recorded by two detectors AB or CD at the same output port of the beam splitter (same-side detection) arise from the bunching effect of HOM interference. On the other hand, the coincidence events recorded by two detectors AC, AD, BC or BD at the opposite output ports of the beam splitter (opposite-side detection) arise from the anti-bunching effect of HOM interference \cite{roger2016coherent}.

\begin{figure*}[htbp]
\centering
\includegraphics[width=1\linewidth]{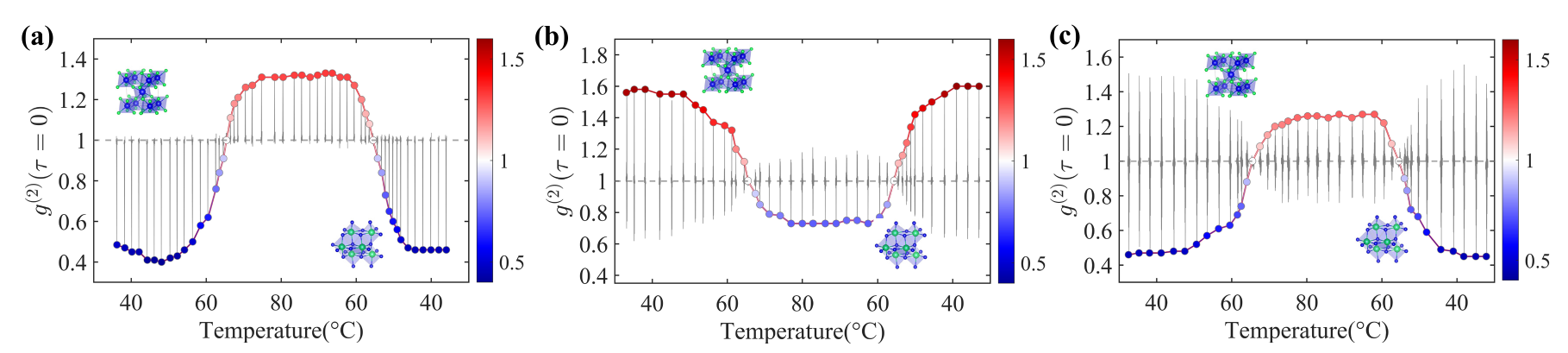}
\vspace{-2mm}
\caption{The measured $g^{(2)}(\tau=0)$ as a function of the heating temperature of VO$_2$ thin film, yielding the dynamical and reversible control of effective quantum photon-photon interaction from bosonic bunching to fermionic antibunching.}
\label{figure_5}
\end{figure*}
Precise scan of the relative time delay enables the direct observation of time-resolved two-photon interference patterns as shown in Fig. \ref{figure_3}. The coincidence probability of bunched photons (labeled by AB and CD) in Fig. \ref{figure_3}(a,k) manifests itself as HOM peaks with average visibility of 71.5\% when the input entanglement is in a symmetric form, but in Fig. \ref{figure_3}(f) as HOM dips with average visibility of 74.5\% when the input entanglement is in an anti-symmetric form. As predicted by $P_{(2,0)}$ and $P_{(0,2)}$ in \eqref{coincidence}, when the VO$_2$ thin film is heating to tune the particle exchange phase response, it is independent of the coincidence probability of bunched photons since the particle exchange phase can be extracted as a global phase in the case that one photon is transmitted and its paired photon is reflected. On the other hand, the coincidence probability of anti-bunched photons (labeled by BC and BD) manifests itself as HOM dips in Fig. \ref{figure_3}(a,k) with average visibility of 57.3\% when the input entanglement is in a symmetric form, but in Fig. \ref{figure_3}(f) as HOM peaks with average visibility of 57.5\% when the input entanglement is in an anti-symmetric form. As predicted by $P_{(1,1)}$ in \eqref{coincidence}, when the VO$_2$ thin film is heating to tune the particle exchange phase response, a phase shift $\pi$ would be introduced into the coincidence probability of anti-bunched photons in the case that both photons are transmitted or reflected. Thus, when the VO$_2$ thin film behaves as a lossless beam splitter, we can observe the conventional HOM interference patterns such that the coincidence probabilities of bunched photons and anti-bunched photons are complementary as a result of energy conservation (Fig. \ref{figure_3}(a,f,k)). However, when the VO$_2$ thin film behaves as a lossy beam splitter by increasing its thermal temperature, the coincidence probability of anti-bunched photons manifests itself as HOM peaks when the input entanglement is in a symmetric form (Fig. \ref{figure_3}(e,o)), but as HOM dips when the input entanglement is in an anti-symmetric form (Fig. \ref{figure_3}(j)), which completely inverts the conventional HOM interference patterns and mimics the fermonic behaviors. The average visibility of anomalous HOM interference in Fig. \ref{figure_3}(e,j,o) can reach 26.5\%, which clearly indicates that the scattering matrix of the VO$_2$ beam splitter becomes a nonunitary operation such that it introduces a phase shift of $\pi$ in the particle exchange phase.

VO$_2$ thin films can be consider as the reconfigurable optical components that are controlled with thermal stimuli in a volatile and reversible fashion, which are written, erased, and rewritten into a refractive-index-changing phase change material with heating and cooling processes. We also experimentally confirm the reversible property of VO$_2$ thin film in dynamically controlling the quantum photon-photon interaction by cooling the heated thin film to a lower temperature. As shown in Fig. \ref{figure_4}, the nonunitary operation can be turned back to the unitary operation, which still enables HOM interference with evident visibility. This (non)unitary transformation has be flexibly switched by dozen of times in our experiment. Therefore, the dynamical control of (non)unitary operations in VO$_2$ thin film are directly presented in the concise yet nontrivial two-photon interference experiment, yielding an obvious transformation between HOM dip and HOM peak in the measured $g^{(2)}(\tau)$ at $\tau=0$ as shown in Fig. \ref{figure_5} (see Supplementary Notes 2 and 5 for more details). 

The coherent absorption of paired photons can be flexibly controlled by harnessing the symmetric properties of incident entanglement as shown in Fig. \ref{figure_3}(e,j,o), where the anti-symmetric property of incident entanglement that makes $\phi_\omega=\pi$ and the lossy beam splitter that makes $\phi_{rt}=\pm \pi$ would result in the coherent perfect absorption of paired photons. This feature may enable the phase change material for quantum state engineering and purification, and may have the potential to control light-matter interactions through quantum interference \cite{Vetlugin2022anti}.

These experimental results agree well with our theoretical prediction, where the slight deviations can be mainly attributed to the imperfect optical components and environmental noise, in particular the different transmission and reflection probabilities reveal that the VO$_2$ thin film is not a perfect balanced beam splitter. This issue may be tackled by using tungsten or nitrogen doped VO$_2$ thin film, or employing VO$_2$ metasurfaces for enhancing the reflective components \cite{liu2025phase}.

\section{Discussion}
Phase change material opens up an alternative degree of freedom for dynamically controlling the photon-photon interaction, and enables the tunable (non)unitary transformation for exploring fascinating physics in quantum optics \cite{bergholtz2021exceptional,benjamin2017science,roger2016coherent}. By introducing the nonunitary transformation matrix, the quantum interference on a lossy beam splitter allows for flexible control over quantum state engineering since the coherent absorption of paired photons can behave as a selective filter by tuning particle exchange phase, or a state distiller by mitigating the mutual interactions between the quantum system and its deleterious environment \cite{mahmoud2025selective}. Strikingly, the dynamical control of photon-photon interaction has great potential in designing novel functions and complex quantum gates in large-scale integrated photonic circuits, in particular for those programmable quantum information tasks such as boson sampling, quantum random walk, and quantum computing \cite{tillmann2013experimental,sansoni2012two}. Loss-induced nonunitary operators can naturally describe the interaction between open physical systems with the environment, and may be connected to non-Hermitian systems that open new avenues for ultrasensitive metrology \cite{xie2024probabilistic,mostafazadeh2002pseudo}. Additionally, VO$_2$ thin film can provide an opportunity for dynamical and continuous control over nonunitary property, which allows for great benefit from the capability to introduce sophisticated nonunitary realization into integrated quantum optics, and paves the way toward large-scale and programmable quantum information platforms \cite{bogaerts2020programmable}. We also envisage that this expansion in quantum interference could inspire new possibilities in other crossover research, such as molecular vibronic spectra in quantum simulation or complex wave functions in quantum chemistry \cite{huh2015boson,aspuru2012photonic}.

\begin{acknowledgments}
This work is supported by the National Natural Science Foundation of China (12034016, 12205107), the National Key R$\&$D Program of China (2023YFA1407200), the Natural Science Foundation of Fujian Province of China (2021J02002), and the program for New Century Excellent Talents in University of China (NCET-13-0495).
\end{acknowledgments}
\bibliography{apssamp}

\end{document}


\setcounter{page}{1}
\begin{center}
\noindent \textbf{\large Supplementary materials for \\
	Dynamical control of quantum photon-photon interaction with phase change material}
\end{center}
\setlength{\parskip}{1\baselineskip}

\noindent \textbf{Supplementary Note 1 \\The phase response limitation of lossy BS}
\setlength{\parskip}{0.35\baselineskip}
   \begin{figure}[!htbp]
 	\centering
 	\includegraphics[width=0.6\linewidth]{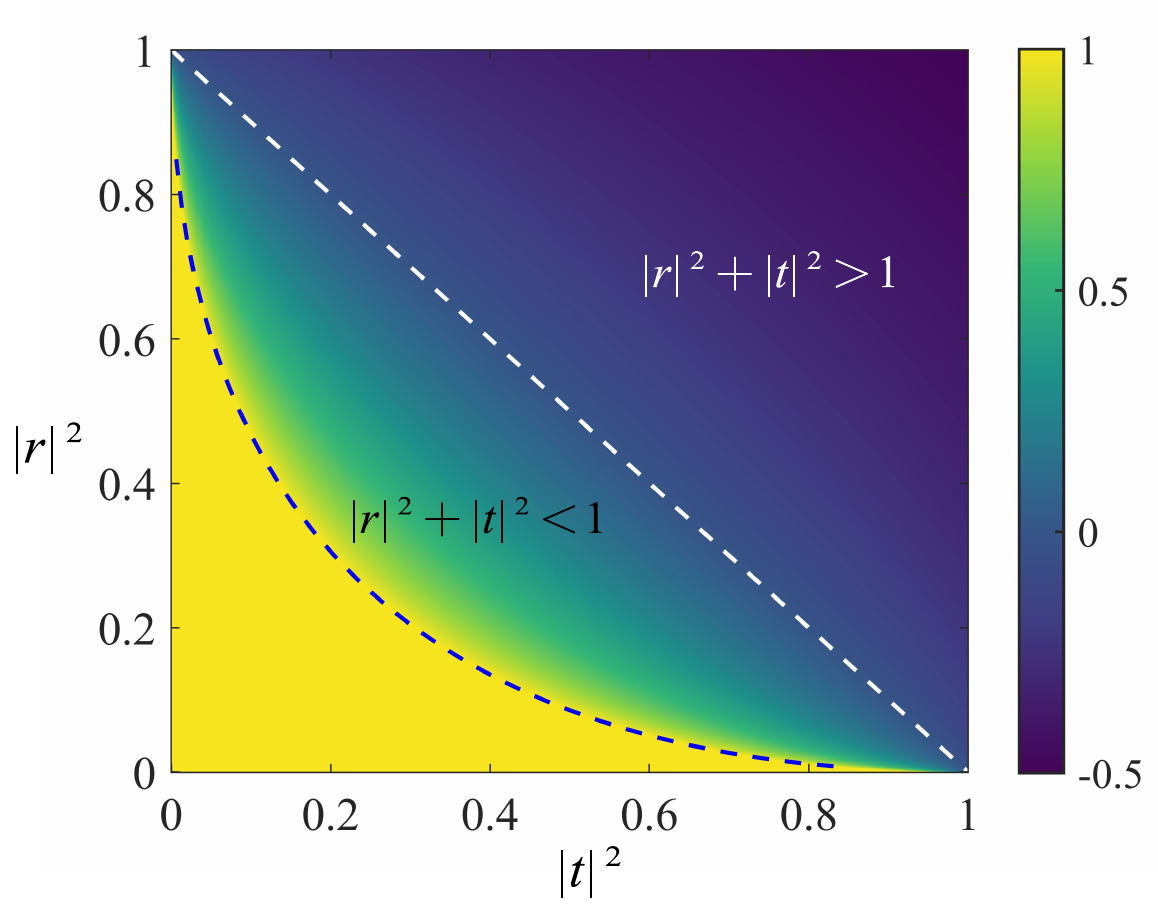}
 	\caption{$|\cos\phi_{rt}|$ as a function of reflectance $|r|^2$ and transmittance $|t|^2$. The white anti-diagonal line $|r|^2+|t|^2=1$ separates the forbidden and allowed regions. The blue dashed line is the curve $(|t|+|r|)^2=1$. The lossy BS that satisfies $(|t|+|r|)^2<1$ allows complete tunability of $\phi_{rt}\in[-\pi,\pi]$.}
 	\label{S2}
 \end{figure}
 
In the conventional BS, the phase response is constrained by the unitary transformation, but it can be broken by a nonunitary BS to obtain different phase responses. Without generality, we consider a symmetric BS with two input ports and two output ports, where the transmission and reflection coefficients are defined as $t=|t|\exp{(i\phi_t)}$ and $r=|r|\exp{(i\phi_r)}$. Since the physical constraint of energy conservation imposes restrictions on the parameters,  the constraints that allow for losses are \cite{Uppu:16,li2021non}
\begin{equation}
	|t|^2+|r|^2\leq1,
\end{equation}
and 
\begin{equation}
	|t\pm r|^2\leq1,
\end{equation}
we can get (Eq.(2) in the main text)
\begin{equation}
	|\cos\phi_{rt}|\leq \frac{1-|t|^2-|r|^2}{2|t||r|},
\end{equation}
 where $\phi_{rt}=\phi_r-\phi_t$ represents the difference between the phase response of the transmitted and reflected photons. Fig. \ref{S2} depicts $|\cos\phi_{rt}|$ as a function of reflectance $|r|^2$ and transmittance $|t|^2$. The anti-diagonal line $|r|^2+|t|^2=1$ separates the allowed region from the forbidden region, which corresponds to the transformation from a lossless BS to a lossy BS. It means that  $|\cos\phi_{rt}|=0$ for the conventional lossless BS, and the correspond phase differences $\phi_{rt}$ must be $\pm\pi/2$. In the region of $|r|^2+|t|^2<1$, $|\cos\phi_{rt}|$ increases with the increasing of loss, the constraints on  phase differences $\phi_{rt}$ gradually relax. The lossy BS that satisfies $(|t|+|r|)^2<1$ allows complete tunability of $\phi_{rt}\in[-\pi,\pi]$. Therefore, it is clear that dynamic control of the phase for the transmission and reflection of two photons can be achieved by modulating the loss of BS. 
 \\
 \\
\noindent \textbf{Supplementary Note 2 \\Two-photon interference based on lossless and lossy beam splitters}
\setlength{\parskip}{0.35\baselineskip}

\begin{figure}[h]
	\centering
	\includegraphics[width=1\linewidth]{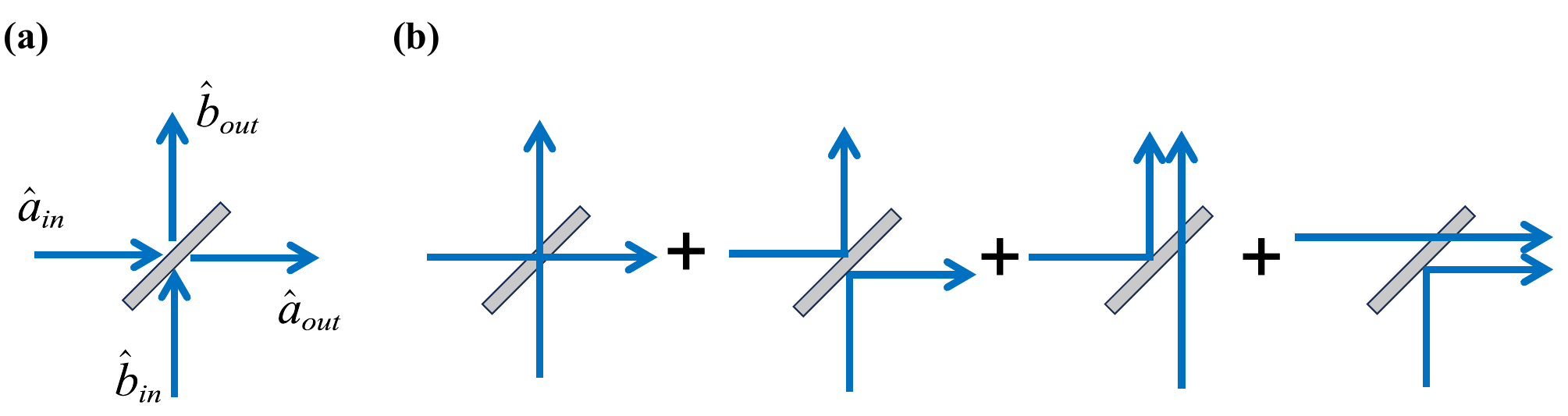}
	\vspace{-2mm}
	\caption{Schematics of (a) a HOM interferometer and (b) the two-particle quantum interference. Two identical particles are simultaneously incident on a beam splitter from opposite input ports $a$ and $b$, which correspond to annihilation operators $\hat{a}_{\mathrm{in}}$ and $\hat{b}_{\mathrm{in}}$, respectively. There are four possible outcomes after interference: two particles are transmitted, two particles are reflected, one particle is transmitted and the other is reflected, and the two-particle state will be a linear superposition of all possibilities. Therefore, the coincidence probabilities of the two particles depend on whether they are bosons or fermions, as well as the phase response at the BS.}
	\label{S1}
\end{figure}
Let us consider a typical configuration of HOM interferometer, where paired particles are incident on a beam splitter from opposite input ports as shown in Fig. \ref{S1}. The paired input and output operators of the BS are related by a scattering transformation as \cite{barnett1998quantum}:
\begin{equation}
	\begin{aligned}
		& \hat{a}_{\mathrm{out}}(\omega)=t\hat{a}_{\mathrm{in}}(\omega)+r \hat{b}_{\mathrm{in}}(\omega)+f_a\hat{n}_a(\omega), \\
		& \hat{b}_{\mathrm{out}}(\omega)=t \hat{b}_{\mathrm{in}}(\omega)+r\hat{a}_{\mathrm{in}}(\omega)+f_b\hat{n}_b(\omega),
	\end{aligned}
\end{equation}
where $\hat{n}_{a,b}$, $\hat{a}_j$ and $\hat{b}_j$ are noise operators and annihilation operators of photons at the input and output ports, respectively. The commutation relations of these operators are satisfied as
\begin{equation}
\begin{aligned}
	&\hat{a}_{j}(\omega)\hat{a}_{j}^{\dagger}(\omega^{\prime})-e^{i\phi}\hat{a}_{j}^{\dagger}(\omega^{\prime})\hat{a}_{j}(\omega)=\delta(\omega_1-\omega_1^{\prime}),\\
	&\hat{b}_{j}(\omega)\hat{b}_{j}^{\dagger}(\omega^{\prime})-e^{i\phi}\hat{b}_{j}^{\dagger}(\omega^{\prime})\hat{b}_{j}(\omega)=\delta(\omega_1-\omega_1^{\prime}),\\
	&\hat{a}_{j}(\omega)\hat{b}_{j}^{\dagger}(\omega^{\prime})-e^{i\phi}\hat{b}_{j}^{\dagger}(\omega^{\prime})\hat{a}_{j}(\omega)=0,\\
	&\hat{b}_{j}(\omega)\hat{a}_{j}^{\dagger}(\omega^{\prime})-e^{i\phi}\hat{a}_{j}^{\dagger}(\omega^{\prime})\hat{b}_{j}(\omega)=0,\\
\end{aligned}
\end{equation}
where $\phi$ ($\phi=0$ for bosons and $\phi=\pi$ for fermions) represents the phase obtained from the exchange of two particles due to quantum statistics.

\begin{figure}[!htbp]
		\centering
		\includegraphics[width=0.6\linewidth]{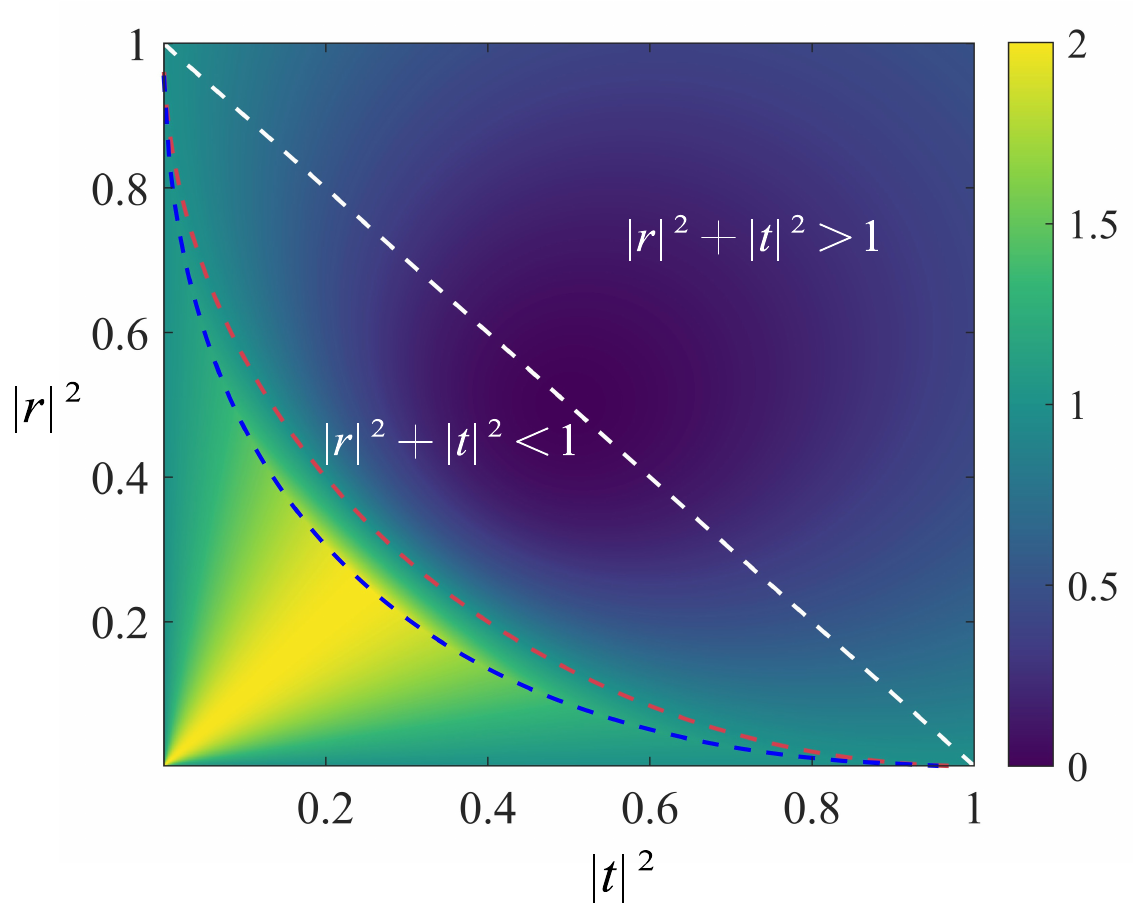}
		\caption{ $g^{(2)}(\tau=0)$ as a function of reflectance $|r|^2$ and transmittance $|t|^2$. The anti-diagonal line $|r|^2+|t|^2=1$ separating the allowed from the forbidden region corresponds to lossless BS. The blue dashed line is the curve $(|t|+|r|)^2=1$, and the red dashed line is the curve $g^{(2)}(\tau=0)=1$.}
		\label{S3}
	\end{figure}

The two-photon state from a spontaneous parametric down-conversion (SPDC) process can be described as \cite{orre2019interference,ramelow2009discrete}
\begin{equation}
|\psi\rangle=\int_0^{\infty} d \omega_s \int_0^{\infty} d \omega_i f\left(\omega_s, \omega_i\right) \hat{a}_{\mathrm{in}}^{\dagger}\left(\omega_s\right) \hat{b}_{\mathrm{in}}^{\dagger}\left(\omega_i\right)|0\rangle,
\end{equation}
where $\omega_s$ and $\omega_i$ are the angular frequency of  signal and idler photons from SPDC, respectively, $f(\omega_s, \omega_i)$ is joint spectral amplitude of the signal and idler photons. The detection field operators of detector 1 (D1) and detector 2 (D2) are defined as
\begin{equation}
	\begin{aligned}
	& \hat{E}_1^{(+)}\left(t_1\right)=\frac{1}{\sqrt{2 \pi}} \int_0^{\infty} d \omega_1 \hat{a}_{\mathrm{out}}\left(\omega_1\right) e^{-i \omega_1 t_1},\\
	&  \hat{E}_2^{(+)}\left(t_2\right)=\frac{1}{\sqrt{2 \pi}} \int_0^{\infty} d \omega_2 \hat{b}_{\mathrm{out}}\left(\omega_2\right) e^{-i \omega_2 t_2}.
    \end{aligned}
\end{equation}
After a delay time $\tau$ on the idler arm and passing through a BS, we have
\begin{equation}
	\begin{aligned}
		& \hat{a}_{\mathrm{out}}(\omega_1)=t \hat{a}_{\mathrm{in}}(\omega_1)+r \hat{b}_{\mathrm{in}}(\omega_1)e^{-i\omega_1\tau}+f_a\hat{n}_a(\omega_1), \\
		& \hat{b}_{\mathrm{out}}(\omega_2)=r \hat{a}_{\mathrm{in}}(\omega_2)+t \hat{b}_{\mathrm{in}}(\omega_2)e^{-i\omega_2\tau}+f_b\hat{n}_b(\omega_2).
	\end{aligned}
\end{equation}
Therefore, the field operators can be rewritten as
\begin{equation}
	\begin{aligned}
		& \hat{E}_1^{(+)}\left(t_1\right)=\frac{1}{\sqrt{2 \pi}} \int_0^{\infty} d \omega_1[t \hat{a}_{\mathrm{in}}(\omega_1)e^{-i \omega_1 t_1}+r \hat{b}_{\mathrm{in}}(\omega_1)e^{-i\omega_1(t_1+\tau)}+f_a\hat{n}_a(\omega_1) e^{-i \omega_1 t_1}],\\
		&  \hat{E}_2^{(+)}\left(t_2\right)=\frac{1}{\sqrt{2 \pi}} \int_0^{\infty} d \omega_2 [r \hat{a}_{\mathrm{in}}(\omega_2)e^{-i \omega_2 t_2}+t \hat{b}_{\mathrm{in}}(\omega_2)e^{-i\omega_2(t_2+\tau)}+f_b\hat{n}_b(\omega_2) e^{-i \omega_2 t_2}].
	\end{aligned}
\end{equation}
The probability $P_{(1,1)}(\tau)$ for coincidence detection of one single photon at each output port can be expressed as
\begin{equation}
	 P_{(1,1)}(\tau)=\iint d t_1 d t_2\langle \psi|\hat{E}_1^{(-)}\hat{E}_2^{(-)}\hat{E}_2^{(+)}\hat{E}_1^{(+)}|\psi\rangle.
\end{equation}
First consider $\hat{E}_2^{(+)}\hat{E}_1^{(+)}|\psi\rangle$, we have
\begin{equation}
\hat{E}_2^{(+)} \hat{E}_1^{(+)}|\psi\rangle=\frac{1}{2 \pi} \int_0^{\infty} \int_0^{\infty} d \omega_1 d \omega_2 e^{-i \omega_1 t_1} e^{-i \omega_2 t_2}\left[r^2f\left(\omega_2, \omega_1\right) e^{-i \omega_1 \tau}+e^{-i\phi}t^2f\left(\omega_1, \omega_2\right) e^{-i \omega_2 \tau}\right]|0\rangle.
\end{equation}
Finally,
\begin{equation}
	\begin{aligned}
		P_{(1,1)}(\tau)= & \iint d t_1 d t_2\langle\psi| \hat{E}_1^{(-)} \hat{E}_2^{(-)} \hat{E}_2^{(+)} \hat{E}_1^{(+)}|\psi\rangle \\
		= &  \int_0^{\infty} \int_0^{\infty} \int_0^{\infty} \int_0^{\infty} d \omega_1 d \omega_2 d \omega_1^{\prime} d \omega_2^{\prime} \delta\left(\omega_1-\omega_1^{\prime}\right) \delta\left(\omega_2-\omega_2^{\prime}\right) \\
		& \times\left[|r|^2e^{-i2\phi_r}f^*\left(\omega_2^{\prime}, \omega_1^{\prime}\right) e^{i \omega_1^{\prime} \tau}+e^{i\phi}|t|^2e^{-i2\phi_t}f^*\left(\omega_1^{\prime}, \omega_2^{\prime}\right) e^{i \omega_2^{\prime} \tau}\right]\\
		&\times\left[|r|^2e^{i2\phi_r}f\left(\omega_2, \omega_1\right) e^{-i \omega_1 \tau}+e^{-i\phi}|t|^2e^{i2\phi_t}f\left(\omega_1, \omega_2\right) e^{-i \omega_2 \tau}\right] \\
		= & \int_0^{\infty} \int_0^{\infty} d \omega_1 d \omega_2\left|\left[|t|^2f\left(\omega_1, \omega_2\right)+e^{i(2\phi_{rt}+\phi)}|r|^2f\left(\omega_2, \omega_1\right) e^{-i\left(\omega_1-\omega_2\right) \tau}\right]\right|^2,
	\end{aligned}
	\end{equation}
we can further simplify this equation to
\begin{equation}
\begin{split}
	P_{(1,1)}(\tau)=&\int_0^{\infty} \int_0^{\infty} d \omega_1 d \omega_2[|t|^4f\left(\omega_1, \omega_2\right)+|r|^4f\left(\omega_2, \omega_1\right)\\
    &+2|t|^2|r|^2f\left(\omega_1, \omega_2\right)f\left(\omega_2, \omega_1\right)\cos(\Delta \tau +2\phi_{rt}+\phi)],
    \end{split}
\end{equation}
where $\Delta=|\omega_2-\omega_1|$ is the difference frequency of two well-separated center frequency bins. 
For a normalized $f\left(\omega_1, \omega_2\right)$, i.e. $\int_0^{\infty} \int_0^{\infty} d \omega_1 d \omega_2|f(\omega_1,\omega_2)|^2=1$,
\begin{equation}
	P_{(1,1)}(\tau)=|t|^4+|r|^4+2|t|^2|r|^2 \int_0^{\infty} \int_0^{\infty} d \omega_1 d \omega_2f\left(\omega_1, \omega_2\right)f\left(\omega_2, \omega_1\right)\cos(\Delta \tau +2\phi_{rt}+\phi).
\end{equation}
If we assume $f\left(\omega_2, \omega_1\right)$ has the exchange symmetry of $f\left(\omega_2, \omega_1\right)=f\left(\omega_1, \omega_2\right)$, we can further simplify the equation as
\begin{equation}
	P_{(1,1)}(\tau)=|t|^4+|r|^4+2|t|^2|r|^2\int_0^{\infty} \int_0^{\infty} d \omega_1 d \omega_2|f\left(\omega_1, \omega_2\right)|^2\cos(\Delta \tau +2\phi_{rt}+\phi).
\end{equation}
Consider two photons with Gaussian spectral amplitude functions, the two-photon coincidence probability can be written as (Eq.(4) in main text)
\begin{equation}
	P_{(1,1)}(\tau)=|t|^4+|r|^4+2|t|^2|r|^2\cos(\Delta \tau +2\phi_{rt}+\phi)\exp{(-\sigma^2\tau^2)},
\end{equation}
 where $\sigma$ is the RMS bandwidth of single photons. When $\tau\rightarrow\infty$, the two photons are completely distinguishable, and the classical probability $P_{(1,1)}(\tau\rightarrow\infty )$ can be calculated as
\begin{equation}
		P_{(1,1)}(\tau\rightarrow\infty )=|t|^4+|r|^4.
\end{equation}
Finally, the second order coherence function $g^{(2)}(\tau=0)$ is
\begin{equation}
	\begin{aligned}
g^{(2)}(\tau=0)&=\frac{P_{(1,1)q}(\tau=0)}{P_{(1,1)}(\tau\rightarrow\infty )}\\
	&=\frac{|t|^4+|r|^4+2|t|^2|r|^2\cos( 2\phi_{rt}+\phi)}{|t|^4+|r|^4}\\
	& =1+\frac{2|t|^2|r|^2\cos( 2\phi_{rt}+\phi)}{|t|^4+|r|^4}.
	\end{aligned}
\end{equation}
Clearly, $g^{(2)}(\tau=0)$ is related to the phase response of BS and the phase generated from the exchange of two particles \cite{benjamin2017anti,vetlugin2022anti,klauck2019observation,friederike2025crossing}. Fig. \ref{S3} depicts $g^{(2)}(\tau=0)$ ($\phi=0$, for bosons) as a function of reflectance $|r|^2$ and transmittance $|t|^2$. In a lossless and balanced BS as $|t|^2=|r|^2=0.5$, the corresponding phase response becomes $\phi_{rt}=\pi/2$. For photons as bosons ($\phi=0$), $g^{(2)}(\tau=0)=0$, which means that two photons entering a balanced lossless BS simultaneously from different ports will be bunched to the same output port. In a lossy and balanced BS as $|t|^2=|r|^2=0.25$ corresponding phase response becomes $\phi_{rt}=\pi$. In this case, for photons as bosons ($\phi=0$), $g^{(2)}(\tau=0)=2$, which means that the two photons will tend to be anti-bunched. Therefore, it is clear that dynamical control of quantum photon-photon interaction can be achieved by modulating the loss of BS.
	
The analogous results are obtained for photons prepared in a symmetric and anti-symmetric entangled state that mimick bosons and fermions: $|\psi\rangle=(|\omega_1\omega_2\rangle+\exp(i\phi_{\omega})|\omega_2\omega_1\rangle)/\sqrt{2}$.
	When photons are simultaneously incident on a lossless BS from different ports in a symmetric entangled state ($\phi_{\omega}=0$), the two photons are bunched. When the photons are in an anti-symmetric state ($\phi_{\omega}=\pi$), the two photons are anti-bunched. On the contrary, the two photons will be anti-bunched when they are simultaneously incident on a lossy BS from different ports in a symmetric entangled state ($\phi_{\omega}=0$), while the two photons are bunched as they are in an anti-symmetric state ($\phi_{\omega}=\pi$).

The two-photons coincidence probability $P_{(2,0)}(\tau)$ or $P_{(0,2)}(\tau)$ detected at the identical output ports can be expressed as
\begin{equation}
	P_{(2,0)}(\tau)=P_{(0,2)}(\tau) =\iint d t_1 d t_2 \langle \psi|\hat{E}_1^{(-)}\hat{E}_1^{(-)}\hat{E}_1^{(+)}\hat{E}_1^{(+)}|\psi\rangle.
\end{equation}
Consider $\hat{E}_1^{(+)}\hat{E}_1^{(+)}|\psi\rangle$, we have
\begin{equation}
	\hat{E}_1^{(+)} \hat{E}_1^{(+)}|\psi\rangle=\frac{1}{2 \pi} \int_0^{\infty} \int_0^{\infty} d \omega_1 d \omega_2 e^{-i \omega_1 t_1} e^{-i \omega_2 t_2}rt\left[f\left(\omega_2, \omega_1\right) e^{-i \omega_1\tau}+f\left(\omega_1, \omega_2\right) e^{-i \omega_2 \tau}\right]|0\rangle.
\end{equation}
Similarly, the probability $P_{(2,0)}(\tau)$ or $P_{(0,2)}(\tau)$ for coincidence detection of two photons at identical output ports can be calculated as
\begin{equation}
	\begin{aligned}
		P_{(2,0)}(\tau)=P_{(0,2)}(\tau)&= \iint d t_1 d t_2 \langle \psi|\hat{E}_1^{(-)}\hat{E}_1^{(-)}\hat{E}_1^{(+)}\hat{E}_1^{(+)}|\psi\rangle \\
		&=|r|^2|t |^2\int_0^{\infty} \int_0^{\infty} d \omega_1 d \omega_2 \left|f\left(\omega_2, \omega_1\right) +e^{i\phi}f\left(\omega_1, \omega_2\right) e^{-i( \omega_1 -\omega_2) \tau}\right|^2,
	\end{aligned}
\end{equation}
we can further simplify the equation as
\begin{equation}
		P_{(2,0)}(\tau)=P_{(0,2)}(\tau)=2|r|^2|t|^2[1+\int_0^{\infty} \int_0^{\infty} d \omega_1 d \omega_2|f\left(\omega_1, \omega_2\right)|^2\cos(\Delta \tau +\phi)].
\end{equation}
Finally, it can be expressed as (Eq.(4) in main text)
\begin{equation}
		P_{(2,0)}(\tau)=P_{(0,2)}(\tau)=2|r|^2|t|^2[1+\cos(\Delta \tau +\phi)e^{-\sigma^2\tau^2}].
\end{equation}
It can be seen that the coincidence probability $P_{(2,0)q}(\tau)$ or $P_{(0,2)q}(\tau)$ is unrelated to the phase response of BS, which means that the bunching and anti-bunching of photons depends on whether they are in a symmetric state or an anti-symmetric state.

\newpage
\noindent \textbf{Supplementary Note 3 \\Characterization of VO$_2$ thin films}
\setlength{\parskip}{0.35\baselineskip}

   \begin{figure}[!htbp]
	\centering
	\includegraphics[width=1\linewidth]{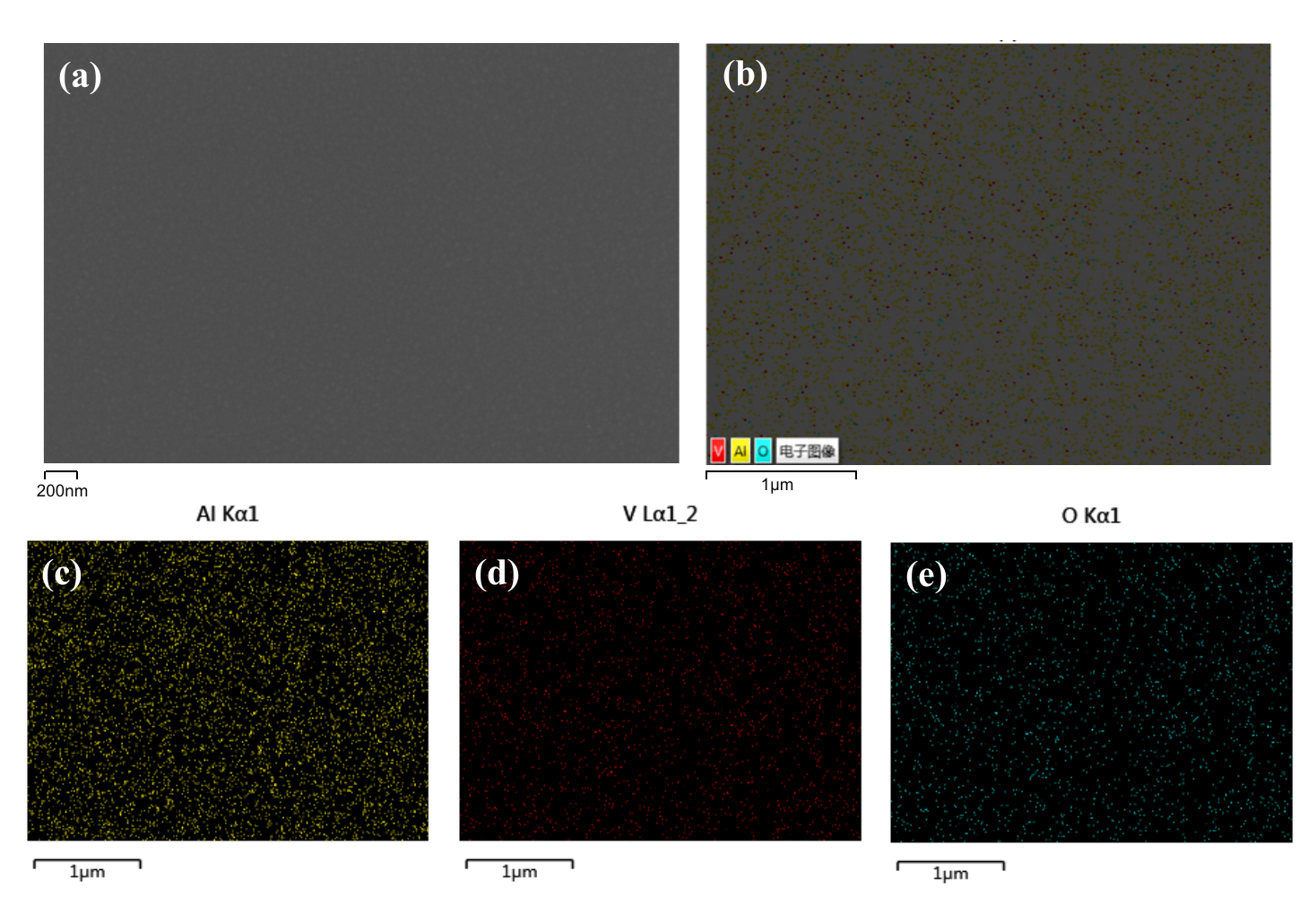}
	\caption{Characterization of the VO$_2$ thin films with thickness of 75nm. (a) A planar-view SEM image of the sample microstructure, captured at a magnification of 30k using a Zeiss SIGMA-HD microscope operating at an accelerating voltage of 8 kV. The image reveals the presence of island-shaped foam-like features on the surface. To further characterize the chemical composition of these structures, energy-dispersive X-ray spectroscopy (EDS) mapping was performed. The corresponding elemental distribution is presented in (b) (overall map), along with individual elemental maps for aluminum (Al), oxygen (O), and vanadium (V) in (c), (d), and (e), respectively.}
	\label{S0}
\end{figure}
The VO$_2$ thin film used in our experiment has been characterized by using SEM and EDS as shown in Fig. \ref{S0}, where the chemical composition figures reveal the fact that the VO$_2$ thin film is deposited on a Al$_2$O$_3$ substrate, and the homogeneity confirms its potential feasibility in controlling the quantum interference in spatial modes.

\newpage
\noindent \textbf{Supplementary Note 4 \\Experimental details}
\setlength{\parskip}{0.35\baselineskip}

In our experiment, we pump two mutually orthogonally oriented 2-mm-long type-II periodically poled potassium titanyl phosphate (PPKTP) crystals with a 405-nm continuous-wave grating-stabilized laser diode. The crystals provide type-II collinear phase matching with the pump (p), signal (s), and idler (i) photons at center wavelengths of $\lambda_p \approx$ 405 nm and $\lambda_{s,i} \approx$ 810 nm at a temperature of 23.5 $^\circ$C. A pair of half-wave and quarter-wave plates in the pump beam are used to set the incident polarization at a diagonal state with a tunable phase shifter as $(\ket{H_p}+\exp(i\phi_\omega)\ket{V_p})/\sqrt{2}$, ensuring that both mutually orthogonal crystals are equally pumped. Consequently, the SPDC process in the first crystal yields the transformation of $\ket{H_p}\rightarrow\ket{H_{\omega_1}V_{\omega_2}}$, and in the second crystal produces the transformation of $\ket{V_p}\rightarrow\ket{V_{\omega_1}H_{\omega_2}}$, where H and V denote horizontal and vertical polarizations, respectively. Followed by a polarization beam splitter that routes orthogonal polarization states into distinct spatial modes, the resultant quantum state can be expressed as $\ket{\Psi}=(\ket{\omega_1\omega_2}+\exp(i\phi_\omega)\ket{\omega_2\omega_1})/\sqrt{2}$.  The frequency detunings $\Delta =|\omega_1-\omega_2|$ and the relative phases $\phi_{\omega}$ can be continuously tuned by adjusting its phase matching temperature \cite{chen2019hong,chen2018polarization}. 

Subsequently, these down-converted photons are coupled into single-mode fibers such that only the Gaussian mode can be transmitted into the VO$_2$ thin film from opposite input ports, which constitutes a HOM interferometry. Additionally, half and quarter wave plates are required to erase the polarization distinguishability. By using a translation stage to vary the relative delay between the signal and idler photons, the resultant non-classical HOM interference patterns can be directly observed. To measure the bunching and anti-bunching probabilities, an in-fiber 50:50 beam splitter is placed at the individual output ports of the beam splitter, which are identified as four single photon detectors labeled A, B, C, and D. The paired photons are detected by silicon avalanche photon diodes, and two-photon coincidence events are identified using a fast electronic AND gate when paired photons arrive at their corresponding detectors within a coincidence time window as $\sim$1ns. Thereinto, the coincidence events recorded by two detectors AB or CD at the same output port of the beam splitter arise from the bunching effect of HOM interference. On the other hand, the coincidence events recorded by two detectors BC or BD at the opposite output ports of the beam splitter arise from the anti-bunching effect of HOM interference. 

\newpage
\noindent \textbf{Supplementary Note 5 \\Experimental measurement of  \bm{$g^{(2)}(\tau=0)$}}
\setlength{\parskip}{0.35\baselineskip}

\begin{figure}[!htbp]
	\centering
	\includegraphics[width=1.0\linewidth]{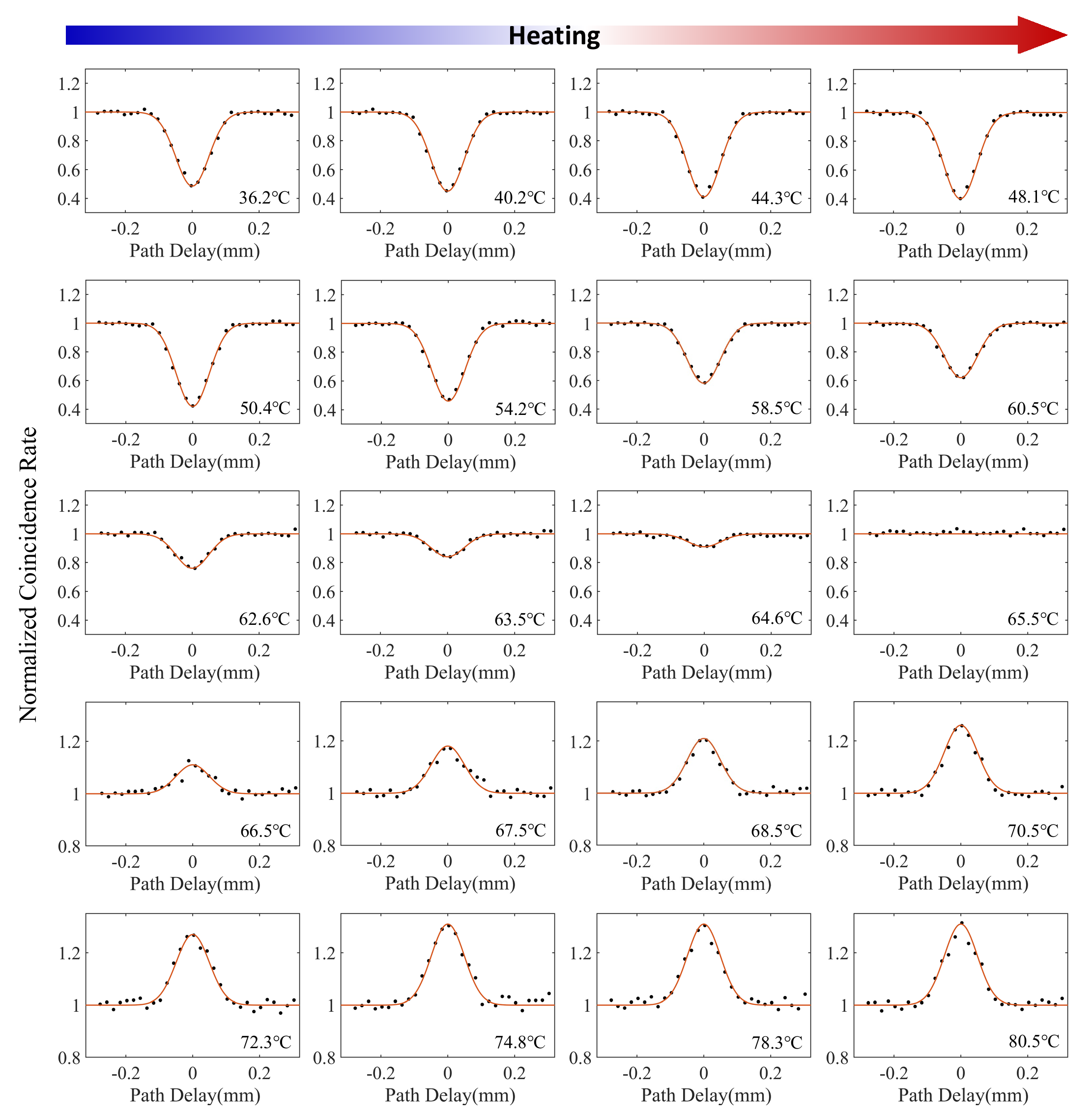}
	\caption{Experimental measurement of HOM interference patterns for symmetric entanglement with degenerate wavelength ($\Delta=0$ THz) when heating the VO$_2$ thin films. }
	\label{S4}
\end{figure}

\begin{figure}[!htbp]
	\centering
	\includegraphics[width=1.0\linewidth]{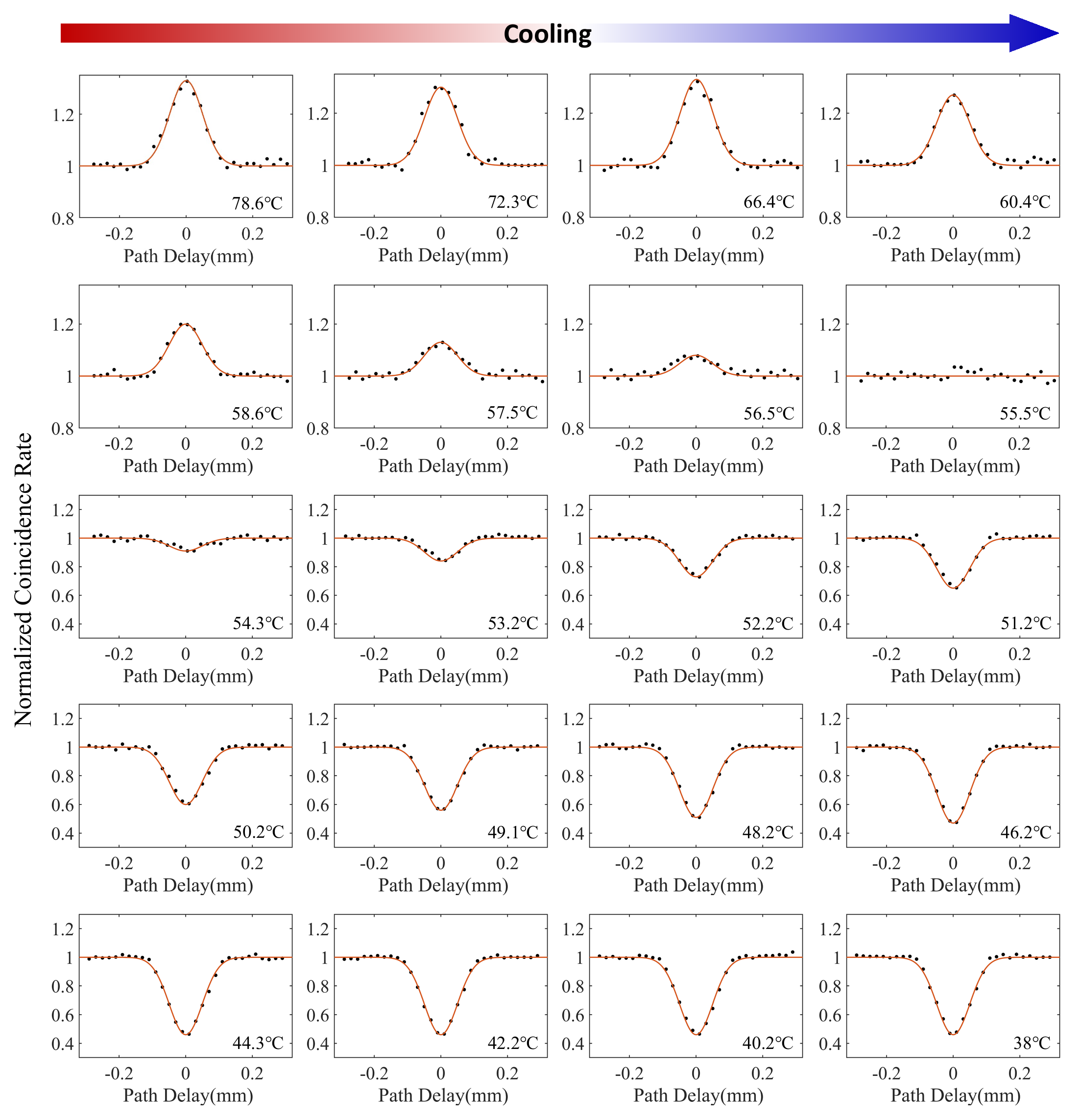}
	\caption{Experimental measurement of HOM interference patterns for symmetric entanglement with degenerate wavelength ($\Delta=0$ THz) when cooling the VO$_2$ thin films. }
	\label{S5}
\end{figure}

\begin{figure}[!htbp]
	\centering
	\includegraphics[width=1.0\linewidth]{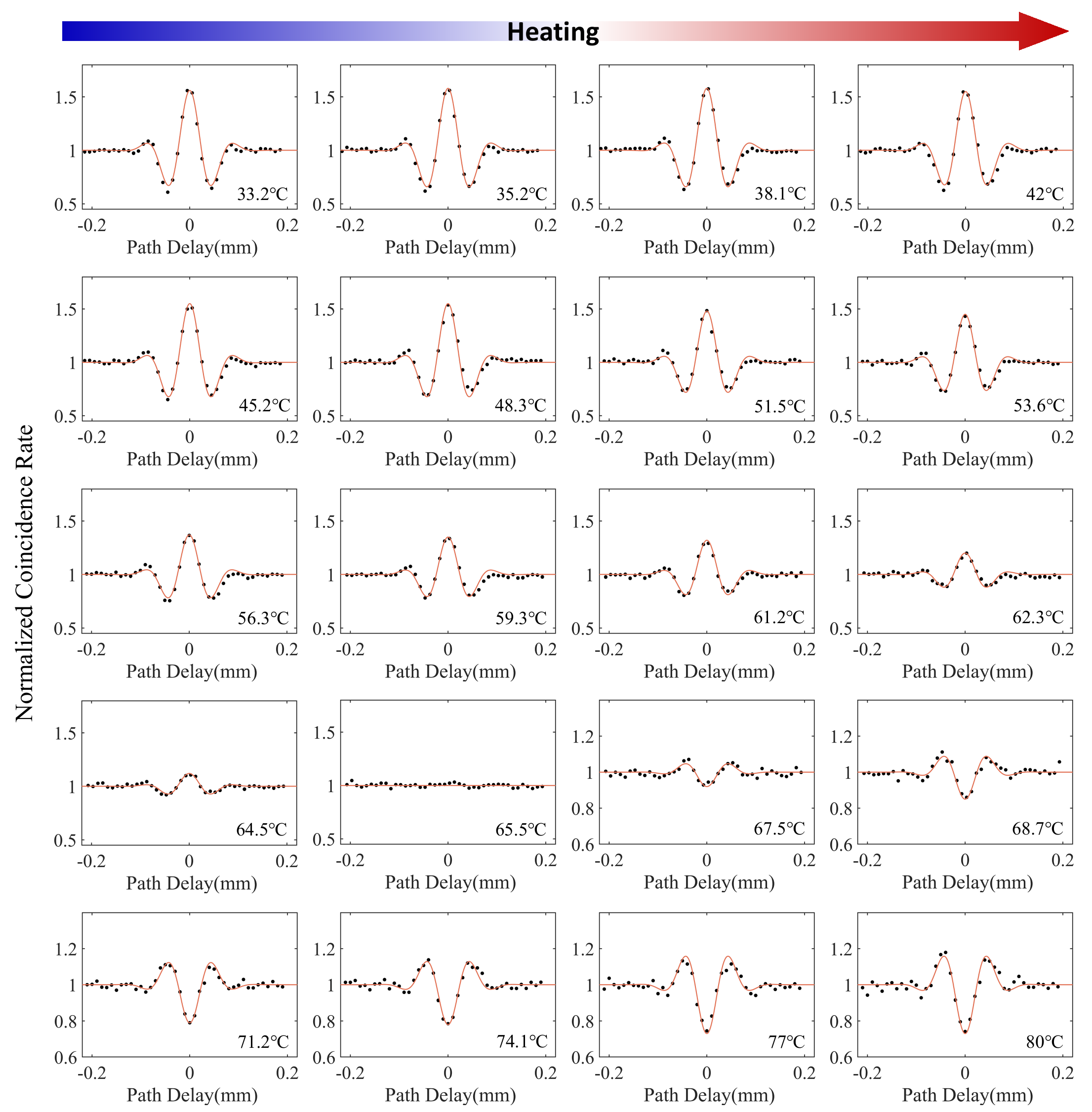}
	\caption{Experimental measurement of HOM interference patterns for anti-symmetric entanglement with non-degenerate wavelengths ($\Delta=2.95$ THz) when heating the VO$_2$ thin films.}
	\label{S6}
\end{figure}

\begin{figure}[!htbp]
	\centering
	\includegraphics[width=1.0\linewidth]{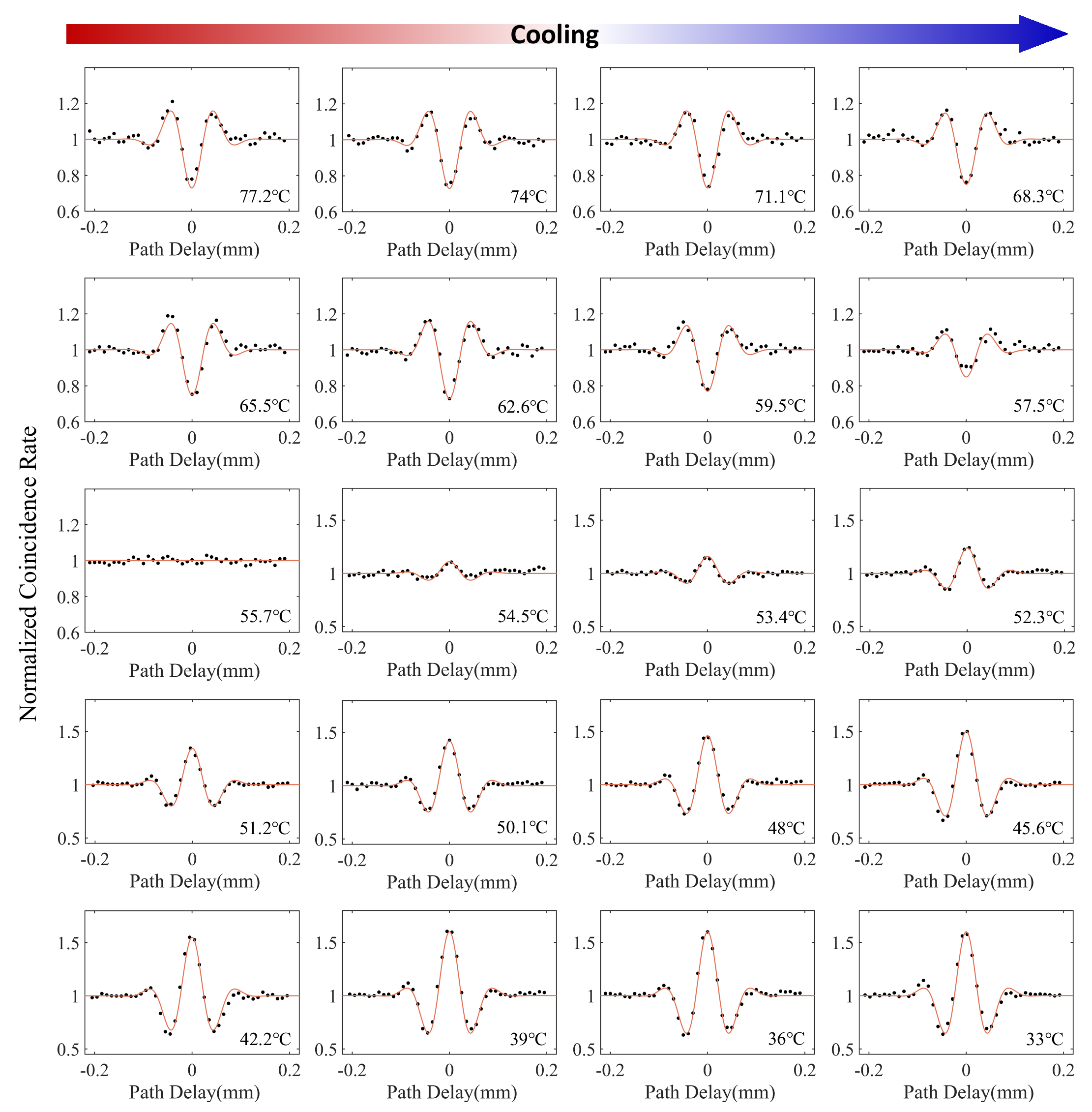}
	\caption{Experimental measurement of HOM interference patterns for anti-symmetric entanglement with non-degenerate wavelengths ($\Delta=2.95$ THz) when cooling the VO$_2$ thin films.}
	\label{S7}
\end{figure}

\begin{figure}[!htbp]
	\centering
	\includegraphics[width=1.0\linewidth]{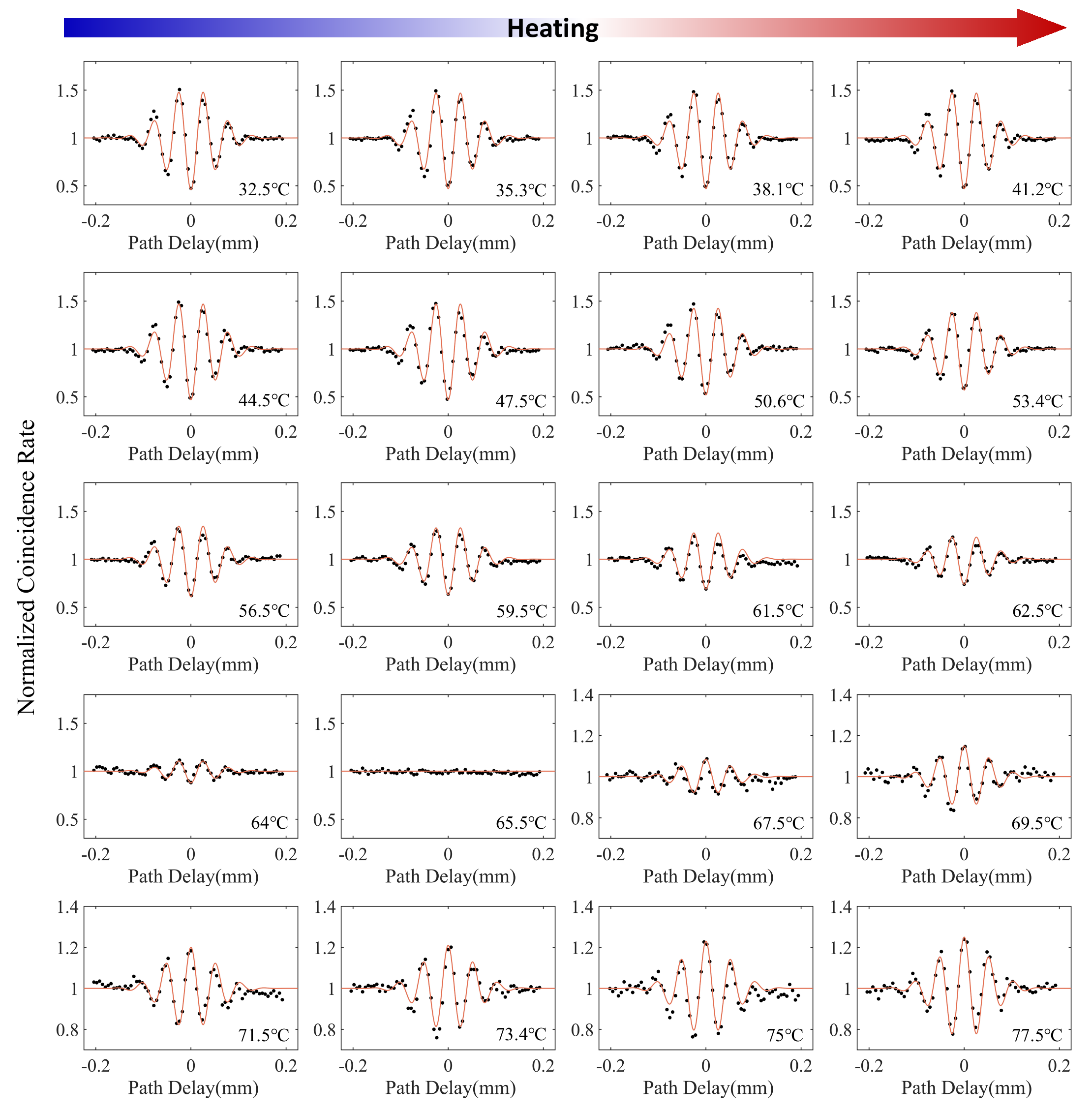}
	\caption{Experimental measurement of HOM interference patterns for symmetric entanglement with non-degenerate wavelengths ($\Delta=5.85$ THz) when heating the VO$_2$ thin films. }
	\label{S8}
\end{figure}

\begin{figure}[!htbp]
	\centering
	\includegraphics[width=1.0\linewidth]{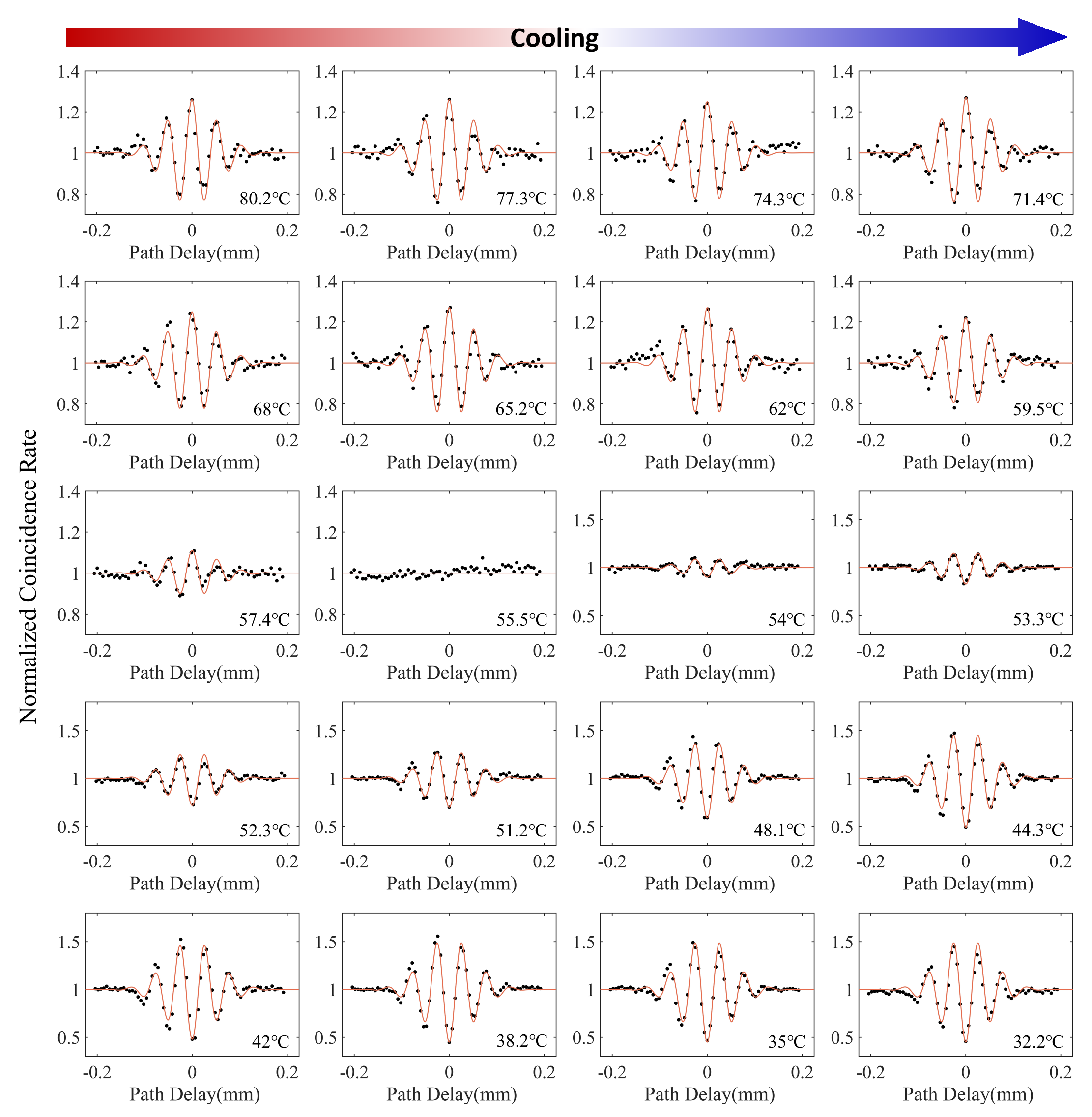}
	\caption{Experimental measurement of HOM interference patterns for symmetric entanglement with non-degenerate wavelengths ($\Delta=5.85$ THz) when cooling the VO$_2$ thin films. }
	\label{S9}
\end{figure}

Fig. \ref{S4} and Fig. \ref{S5} show the HOM interference patterns for measuring the $g^{(2)}(\tau=0)$ of symmetric entanglement with degenerate wavelength ($\Delta=0$ THz). Fig. \ref{S6} and Fig. \ref{S7} show the HOM interference patterns for measuring the $g^{(2)}(\tau=0)$ of  anti-symmetric entanglement with non-degenerate wavelengths ($\Delta=2.95$ THz) . Fig. \ref{S8} and Fig. \ref{S9} show the HOM interference patterns for measuring the $g^{(2)}(\tau=0)$ of  symmetric entanglement with non-degenerate wavelengths ($\Delta=5.85$ THz).

\newpage
\newpage
\newpage
\newpage
\newpage
\newpage

\newpage
\bibliography{Sup.bib}